\documentstyle[emulateapj,psfig]{article}

\def\lta{\mathrel{\spose{\lower 3pt\hbox{$\mathchar"218$}}
     \raise 2.0pt\hbox{$\mathchar"13C$}}}
\def\gta{\mathrel{\spose{\lower 3pt\hbox{$\mathchar"218$}}
     \raise 2.0pt\hbox{$\mathchar"13E$}}}

\def\mathnew{\mathsurround=0pt}

\def\simov#1#2{\lower .5pt\vbox{\baselineskip0pt \lineskip-.5pt
\ialign{$\mathnew#1\hfil##\hfil$\crcr#2\crcr\sim\crcr}}}

\begin{document}

\title{Pseudo-Newtonian Potentials to Describe the Temporal Effects on Relativistic 
Accretion Disks around Rotating Black Holes and Neutron Stars}

\author{ Banibrata Mukhopadhyay \and Ranjeev Misra}

\affil{Inter-University Center for Astronomy and Astrophysics, Post Bag 4,
Ganeshkhind, Pune-411007, India; bm@iucaa.ernet.in; rmisra@iucaa.ernet.in}

\begin{abstract}

Two pseudo-Newtonian potentials, which approximate the 
angular and epicyclic frequencies of the relativistic accretion disk around
rotating (and counter rotating) compact objects, are presented. One of them, the Logarithmically Modified Potential,
is a better approximation for the frequencies while the other, the Second-order
Expanded potential, also reproduces the specific energy for circular orbits in close
agreement with the General Relativistic values. These potentials may be included in time 
dependent hydrodynamical simulations to study the temporal behavior of such 
accretion disks.  

\end{abstract}


\section{Introduction} \label{sec: I}
X-ray binaries are known to be powered by accretion disks around
Neutron Stars and Black Holes. The rapid variability of these sources
indicate that the X-ray emission arises from the inner accretion disk
where the effects of strong gravity are important. High frequency
($\approx$ kHz) Quasi Periodic Oscillations (QPO) have been observed in 
neutron star systems (see van~der~Klis~2000 for a review) while  slightly lower
frequencies ($\approx 450$ Hz) QPOs have been detected in
black hole systems (Strohmayer 2001). For neutron star systems the kHz QPO
tends to be observed in pairs. Associating these frequencies  
with a Keplerian frequency in the disk, leads to the conclusion that
the phenomena originate at radii less than 20 gravitational radius ($r_g \equiv GM/c^2$). 

A number of theoretical ideas have been proposed to explain the phenomenology 
of kilohertz QPO.  In all these models, one of the two frequencies observed 
in neutron star systems, is identified 
as the Keplerian frequency of the innermost orbit of an accretion disk. The sonic point
model (Miller et al.~1998) identifies the second frequency as 
the beat of the primary QPO with the spin of the neutron star, while
according to the two oscillators model (Osherovich \& Titarchuk~1999), the secondary 
frequency is due to the transformation of the primary (Keplerian) 
frequency in the rotating frame of the neutron star magnetosphere.
On the other hand, Stella \& Vietri 
(1999) have proposed a General Relativistic (GR) precession/apsidal motion model, 
wherein the primary frequency is the Keplerian frequency of a slightly eccentric
orbit and the secondary is due to the relativistic apsidal motion of this
orbit i.e. the secondary frequency is the Keplerian frequency minus the
epicyclic one. These models in general are based on identifying the characteristic
frequencies of the system with observed ones and often do not address the issue
of how such oscillations occur in the accreting flow. 

A complete understanding
of the QPO phenomena would require a self consistent hydrodynamical simulation
of the accreting flow in general relativity.  While such an ambitious endeavor
has been impeded for several reasons, the main difficulties can be identified
to be (a) the development of a self-consistent turbulent viscosity and (b)
the inclusion of GR effects. In hydrodynamical simulations,
turbulent viscosity has typically been introduced in a parametric
form like the $\alpha$-parameterization (e.g. Taam \& Lin 1984). Since 
the temporal behavior of accretion disks is expected to depend on the form of the viscosity 
law, the results of such simulations were not conclusive. 
A promising mechanism for driving the turbulence responsible for angular 
momentum and energy transport is the action of the magneto-rotational 
instability (MRI) that is expected to take
place in such disks (Balbus \& Hawley 1991). Recent 3D magneto hydrodynamical
(MHD) simulations have shown that indeed the MRI can give rise to a turbulent
viscosity which leads to the accretion flow in a Keplerian disk (Hawley, 
Balbus \& Stone 2001). While presently such simulations do not include 
radiation (and hence do not describe optically thick accretion flow), it is
expected that self-consistent simulations will be possible 
in the near future and the
temporal behavior of accretion disks can be studied with confidence.
 
Despite these recent advances, it is still extremely difficult to
simulate realistic accretion flows in a complete GR
framework. However, relativistic effects may be approximately simulated
by using modified Newtonian (or Pseudo-Newtonian) potentials
in the non-relativistic radial-momentum equation. 
Paczy\'nski \& Wiita (1980) proposed such a pseudo-Newtonian potential
which has been frequently used in simulations (e.g. Milsom \& Taam 1997;
Hawley \& Balbus 2002). Here the Newtonian potential has been replaced by 
$\phi = GM/(r-2r_g)$. The
attractive feature of the potential is that it reproduces the
last stable orbit exactly and the specific energies
of circular orbits within $10\%$ of the GR values
(i.e. for Schwarzschild geometry). Several other pseudo-Newtonian potentials
have been proposed and used in the literature (e.g. Chakrabarti \& Khanna 1992).
Artemova et al. (1996) have considered several such potentials and concluded
that the Paczy\'nski-Wiita potential is better than the rest based on the
above criteria for non-rotating compact objects. 
Recently, Mukhopadhyay (2002) has proposed a pseudo-
potential which is valid for rotating compact objects. This potential
reproduces the GR values of last stable orbit exactly and is a 
good approximation ($< 10\%$ error) for the specific energy at last stable circular orbit 
in case of Kerr geometry. It also reduces to the Paczy\'nski-Wiita potential when the spin 
of the black hole is set to zero.

However, these potentials are not a good  
approximation (with error $ > 50\%$) for the angular and epicyclic 
frequencies for radii $ < 20 r_g$. Thus, while they are
adequate to approximate the relativistic effects for
a steady state accretion disk, they can not quantitatively reproduce
the temporal behavior of a disk since that is expected to depend on the 
disk's characteristic
(i.e. the angular and epicyclic) frequencies. Nowak \& Wagoner (1991) have
proposed a potential for a non rotating black hole which reproduces 
the Keplerian frequencies (with deviations $<15\%$) and the epicyclic
frequencies (with deviations less than $45\%$) and hence is better than
the Paczy\'nski-Wiita potential for such applications.     

In this paper, we present two pseudo-Newtonian potentials which may
be used to simulate the relativistic time varying effects in accretion
disks around a compact object that may be 
co-rotating or counter-rotating with respect to the disk
with the  spin parameter $a < 0.99$. For faster spin rates 
the predictions of these potentials deviate pronouncedly
(with errors $> 200\%$) and hence are no longer a good approximation. 
The first has been named the {\it Second-order 
Expansion Potential (SEP)} since it contains terms up to $(r_{ms}/r)^2$, where
$r_{ms}$ is the marginally stable orbit. This potential reproduces
the specific energy  and the angular frequency with deviations $<10\%$ and 
$< 25\%$, respectively from GR values (i.e. for Kerr geometry). The deviations
in epicyclic frequency range from $25-170\%$ (for $a\leq 0.9$) 
depending on the spin rate
of the compact object. When the object is not rotating, the potential
reduces to the one proposed by Nowak \& Wagoner (1991). The second
has been named {\it Logarithmically Modified Potential (LMP)} since 
it contains a logarithmic term. This potential reproduces well 
the angular (with deviations $< 20\%$ for co-rotating and $<40\%$ for counter-rotating flows) 
and epicyclic frequencies (with deviations $< 60\%$) but predicts specific energies which are around 
$30\%$ different from the GR values.

\section{Pseudo-Newtonian Potentials} \label{sec: II}

Since hydrodynamical code directly require the gravitational acceleration, it is
practical to modify the Newtonian force instead of the potential. In
terms of such a modified force per unit mass ($F$), the angular ($\Omega$) 
and epicyclic ($\kappa$) frequencies
are given by
\begin{equation}
\Omega^2 = \frac{F}{R} 
\end{equation}
and
\begin{equation}
\kappa^2 = {2 \Omega \over R} \frac{d}{dR} (\Omega R^2) = {1 \over R^3}\frac{d}{dR}(FR^3).
\end{equation}
These Newtonian (or pseudo-Newtonian) frequencies should match with the GR
ones ($\Omega_{GR}$ and $\kappa_{GR}$) as seen by an observer at infinity. 
In terms of the dimensionless radial coordinate ($r = R/r_g$) and spin
parameter ($a$) these frequencies are given by (e.g. Semer\'ak \& Z\'acek 2000),
\begin{equation}
\Omega_{GR} = {1 \over r^{3/2}+a}
\end{equation}
and
\begin{equation}
\kappa_{GR}^2 = \left({\Omega_{GR} \over r}\right)^2[ \Delta - 4(\sqrt{r} -a)^2]
\end{equation}
where $\Delta = r^2 - 2r+a^2$. For the above equations and rest of the paper 
we have used dimensionless quantities by setting $G$, $M$ and $c$ to be unity.
 
The modified force should also reproduce the marginal stable
radius ($r_{ms}$) given by (Bardeen 1973)
\begin{eqnarray}
r_{ms} & = &3 + Z_2 \mp[(3-Z_1)(3+Z_1+2Z_2)]^{1/2} \\
\nonumber
Z_1 & = & 1 +(1-a^2)^{1/3}[(1+a)^{1/3}+(1-a)^{1/3}] \\
\nonumber
Z_2 & = & (3a^2+Z_1^2)^{1/2}, 
\end{eqnarray}
where the '-' ('+') sign is for the co-rotating (counter-rotating) flow.

Since either one of the relativistic frequencies 
($\Omega_{GR}$ and $\kappa_{GR}$) can specify the form of the required
modified force, $F$ (from Eqns. 1 or 2), it is clear that a modified
Newtonian force cannot reproduce both the frequencies exactly. Hence
an appropriately chosen modified force should correspond to frequencies which
have minimal deviation from the GR values. Here we
present two such modified (or pseudo-Newtonian) forces. Both the potentials
are constructed in such a manner that the marginally stable orbit ($r_{ms}$) is always
same as the relativistic value (i.e. in  Kerr geometry).

\subsection{ Second-order Expansion Potential (SEP)}

For this potential the Newtonian force per unit mass is modified to be,
\begin{equation}
F = \Omega^2 r = {1\over r^2}\left[ 1 - \left({r_{ms}\over r}\right) + \left({r_{ms}\over 
r}\right)^2\right]
\end{equation}
where $r_{ms}$ is given by Eqn. (5). The corresponding epicyclic frequency
is
\begin{equation}
\kappa = {1\over r^{3/2}}\left[1-\left(\frac{r_{ms}}{r}\right)^2\right]^{1/2}.
\end{equation}

In Figs. 1, 2, and 3, the variation of angular and epicyclic 
frequencies with radii are compared with GR values for
three different values of the spin parameters $a = 0, 0.5, 0.9$. 
Figure 4 shows the variations of these frequencies 
for spin parameter $a = 0.99$ where the deviations from the GR values are 
large and the potentials described in the work are no longer a good approximation,
particularly for the epicyclic frequency.
The main advantage of this potential is its relative simplicity and that the
angular frequencies deviate from the GR values by
less than or equals to $25\%$. The specific energy (i.e. the energy per unit mass for
a circular orbit) is also close (error is at most $\sim 10\%$ for all values of the Kerr parameter
including $a=1$) to the relativistic values (see Fig. 5). Its disadvantage is that $\kappa$ deviates 
from $\kappa_{GR}$
by around $40\%$ for low and by nearly $150\%$ for high spin values ($a \approx 0.9$) 
of the compact object (Fig. 3). However, for higher counter-rotation of the compact
object the error in $\kappa$ reduces to $\sim 25\%$.

\subsection{ Logarithmically Modified Potential (LMP)}

Here the Newtonian force is modified to be
\begin{equation}
\nonumber
F = {1\over r^2}\left[1 + r_{ms}\left\{{9\over 20}{(r_{ms}-1) \over r} - {3 \over 2 r} 
log \left(r \over (3 r - r_{ms})^{2/9}\right)\right\}\right]\\
\end{equation}
and the corresponding epicyclic frequency is given by
\begin{equation}
\kappa^2 = {3 \over 2 r^4} {(r-r_{ms})(2 r -r_{ms}) \over (3 r - r_{ms})}.
\end{equation}
Note that the term in the force which depends on $r^{-3}$ does not contribute
to $\kappa$ (Eqn. 2). The logarithmic form of the modified force (Eqn. (8))
was obtained by integrating the epicyclic frequency expression (9) whose form
was guessed to be a good approximation. 
The advantage of this potential is that both the
angular and epicyclic frequencies are generally better comparable with the GR
values than the SEP (Figs. 1, 2 and 3). Its disadvantage is that the specific
energy deviates by more than $30\%$ from the GR values 
which is substantially larger than the deviation for SEP (Fig. 5).

\section{Summary and Discussion} \label{sec: III}

In this work, we have presented two pseudo-Newtonian potentials which
approximate the general relativistic effects on an accretion disk 
around rotating compact objects. These two potentials are designed particularly 
to approximate the angular and epicyclic frequencies of
the accretion disk as seen by an observer at infinity. Table 1 summarizes
the results by comparing the maximum percentage deviations from 
relativistic values (in Kerr geometry)
for the two potentials and comparing them with those of another standard
pseudo-potential. 

The SEP not only approximates the frequencies well,
but also the specific energies for circular orbits turn out to be 
remarkably close to the
relativistic values. Thus based on such criteria, this potential is
better than other pseudo-Newtonian potentials given in the literature
and can be used to simulate both the steady state and time varying accretion disks.
The LMP while being a better approximation to the frequencies than SEP, gives rather large
($\approx 30\%$) deviation from the GR results for the specific energies. 
Hence its utility is perhaps limited to the time-dependent studies of accretion disks. 

Which one of these two potentials should be used in a hydrodynamical simulation
depends on problem being addressed. Acoustic waves (which depend
on the epicyclic frequencies) would perhaps be better simulated by
the LMP while the SEP may be more suited for the long term temporal behavior 
(which may depend also on the energy dissipation). Moreover, a temporal behavior
detected in a simulation could be an artifact of the pseudo-Newtonian
potential rather than true GR effects. Hence, it will
be prudent to confirm the behavior using both the potentials. 
Since the mathematical forms of the two potentials are quite different
any temporal behavior detected for both the potentials would imply
that the behavior is indeed due to relativistic effects. Use of these
potentials in hydrodynamical simulations of accretion disk will help
in the understanding of relativistic effects and may serve as a 
guideline for advanced simulations in general relativity.

\acknowledgements


\begin{figure}
\epsscale{.60}
\plotone{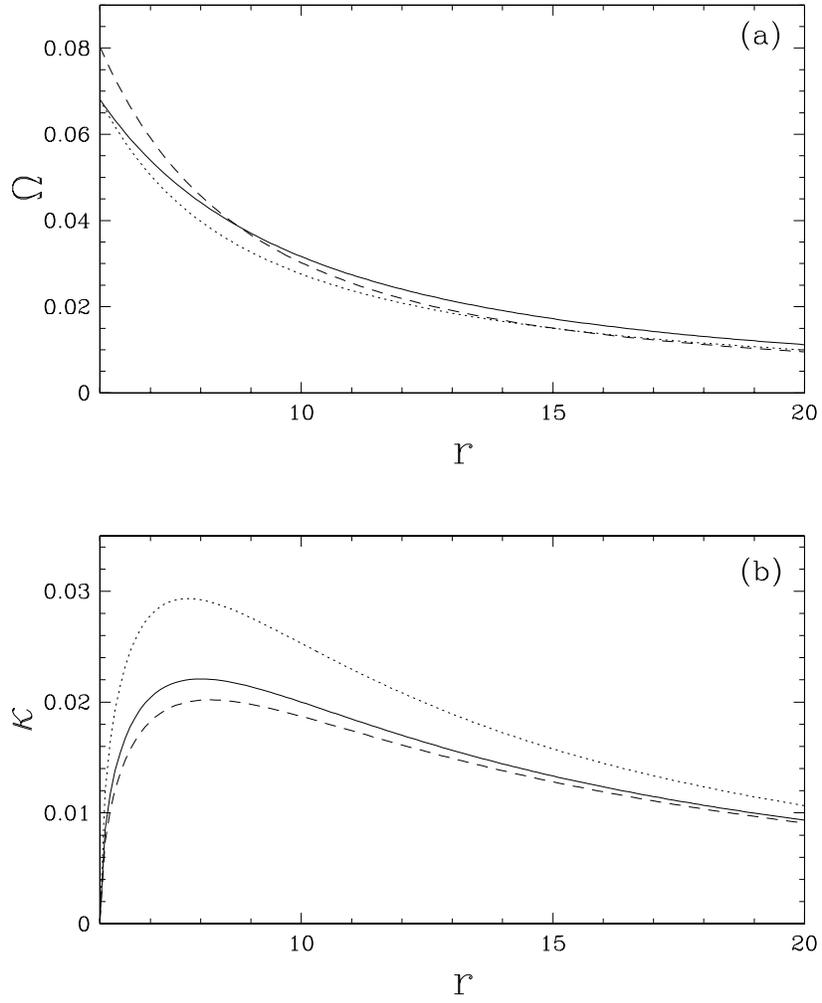}
\caption{Variation of (a) angular and (b) epicyclic frequencies with radii for a
non rotating compact object ($a = 0$). The solid line is for  general
relativity, dotted line is for the SEP (in this case same as the potential given by 
Nowak \& Wagoner 1991) and the dashed line is for the LMP.
\label{fig1}}
\end{figure}

\begin{figure}
\epsscale{.60}
\plotone{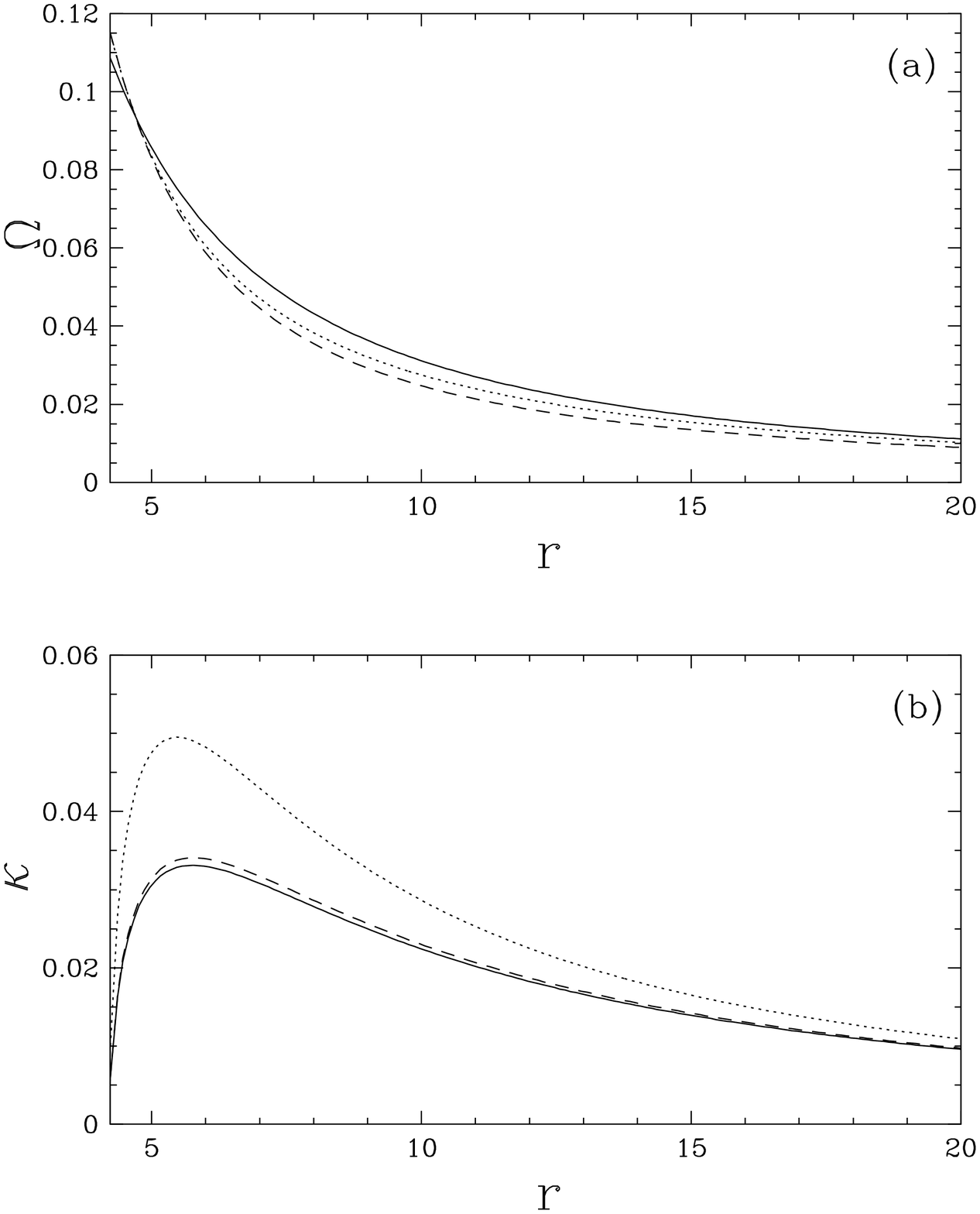}
\caption{Same as in Fig. 1 except that $a = 0.5$ 
\label{fig2}}
\end{figure}

\begin{figure}
\epsscale{.60}
\plotone{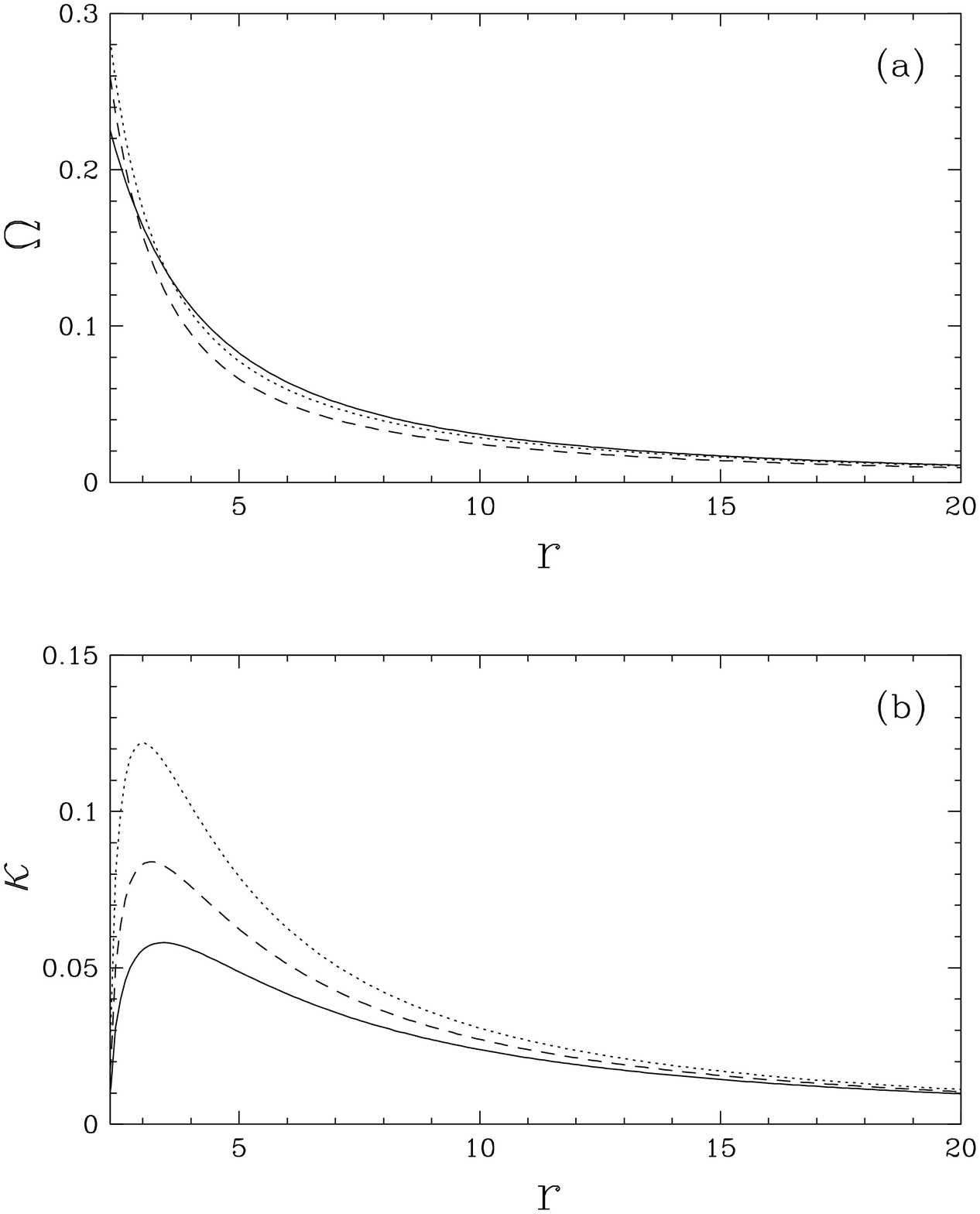}
\caption{Same as in Fig. 1 except that $a = 0.9$.
\label{fig3}}
\end{figure}

\begin{figure}
\epsscale{.60}
\plotone{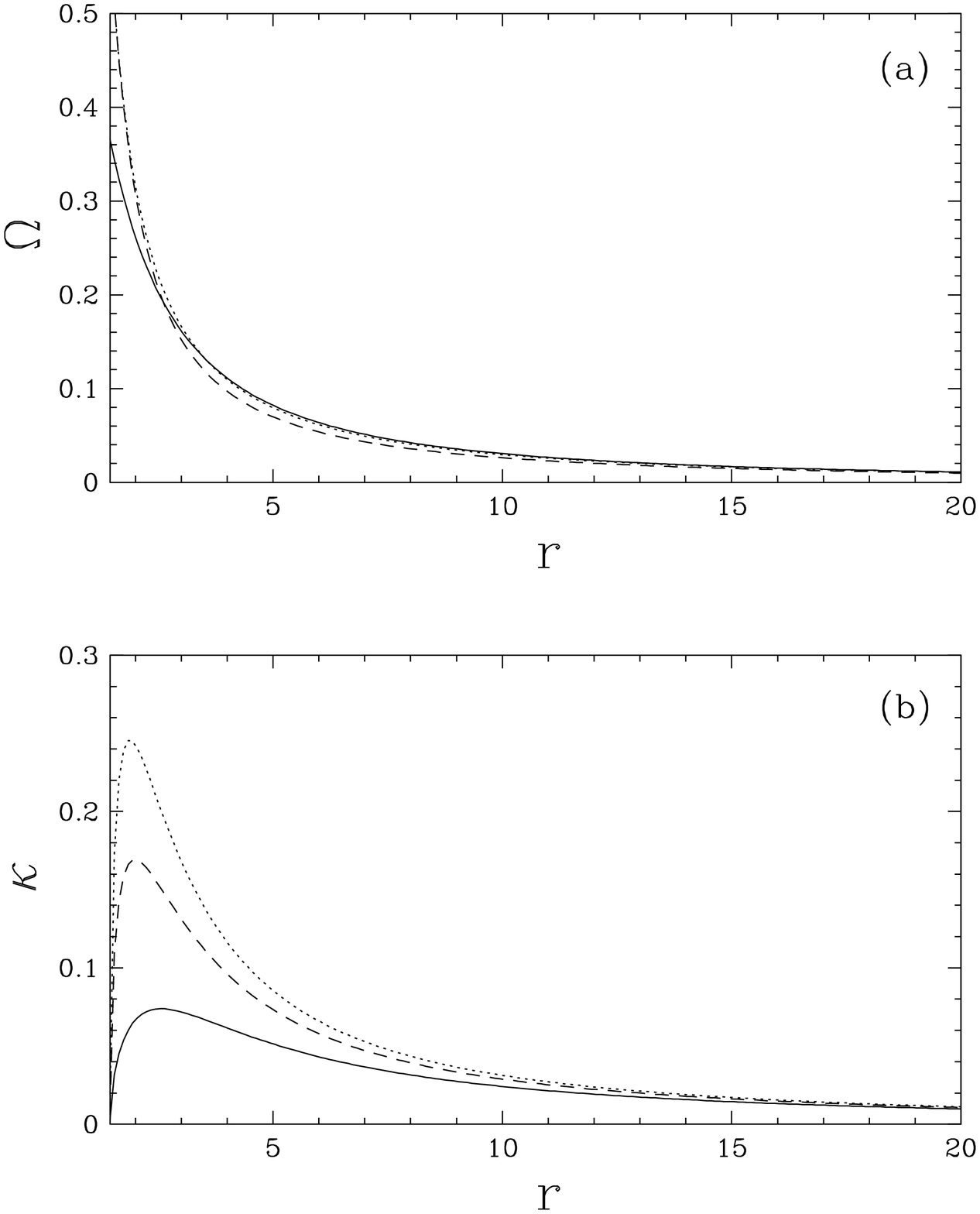}
\caption{Same as in Fig. 1 except that $a = 0.99$.
\label{fig3}}
\end{figure}

\begin{figure}
\epsscale{.60}
\plotone{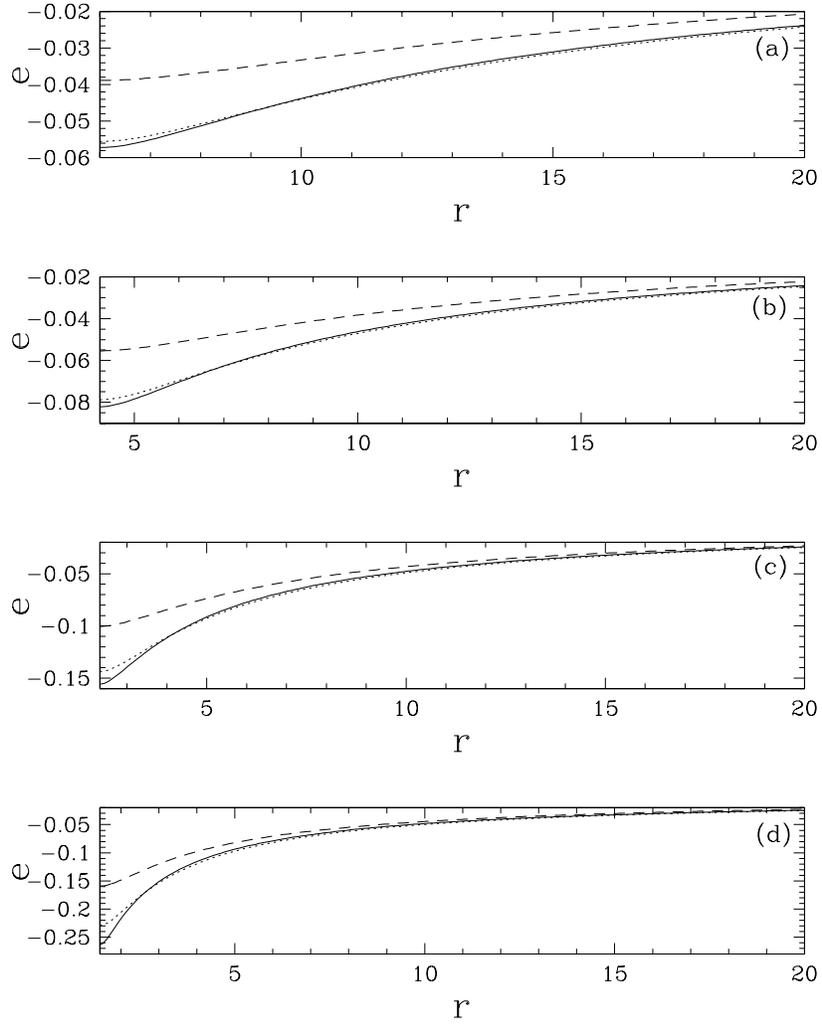}
\caption{Variation of specific energy for circular orbits with radii for 
(a) non-rotating compact object as $a = 0$, (b) rotating compact object as $a = 0.5$,
(c) rotating compact object as $a = 0.9$ and (d) rotating compact object as $a = 0.99$. 
The solid line is for  general relativity, dotted line is for the SEP while 
the dashed line is for the LMP.
\label{fig4}}
\end{figure}

\begin{table*}
{\centerline{\Large Table 1}}
\vspace{0.2cm}
\noindent
Maximum percentage deviation of angular frequency ($\Delta \Omega$),
epicyclic frequency ($\Delta \kappa$) and specific energy ($\Delta e$) for
the Second-order Expanded Potential (SEP), Logarithmically Modified Potential (LMP) and
the potential given by Mukhopadhyay (2002) (M). $a$ is the spin parameter with positive
sign indicating co-rotation and negative sign indicating counter-rotation and
$r_{ms}$ is the radius of the marginal stable orbit.
\begin{center}

\begin{tabular}{cccccc}
\tableline
\tableline
Pseudo-Newtonian Potential & $a$  & $\Delta \Omega$ (\%)  & $\Delta \kappa$ (\%) & $\Delta e$ (\%) & $r_{ms}$\\
\tableline
\tableline
SEP & 0.99 & 57 & 445 &  13 & 1.45 \\

LMP & 0.99 & 57 &  240 & 40 & 1.45 \\

M   & 0.99 & 180 & 800  & 14  & 1.45 \\
\tableline
SEP & 0.9 & 25 & 168 &  8 & 2.32 \\

LMP & 0.9 & 22 &  64 & 35 & 2.32 \\

M   & 0.9 & 100 & 300  & 12 & 2.32 \\
\tableline
SEP & 0.5 & 12 & 65 &  4 & 4.23 \\
LMP & 0.5 & 22 & 17 & 32 & 4.23 \\
M   & 0.5 & 63 & 122  & 10 & 4.23 \\
\tableline
SEP\tablenotemark{*} & 0.0 & 13 & 42 &  3 & 6.00 \\
LMP & 0.0 & 18 & 13 & 32 & 6.00 \\
M\tablenotemark{**}   & 0.0 & 50 & 84  & 9 & 6.00 \\
\tableline
SEP & -0.5 & 14 & 31 &  2 & 7.55 \\
LMP & -0.5 & 31 &  20 & 31 & 7.55 \\
M   & -0.5 & 44 & 67  & 9 & 7.55 \\
\tableline
SEP & -0.9 & 14 & 26 &  2 & 8.72 \\
LMP & -0.9 & 42 & 23 & 31 & 8.72 \\
M   & -0.9 & 42 & 60  & 9 & 8.72 \\
\tableline
SEP & -0.99 & 15 & 25 & 2  & 8.97 \\
LMP & -0.99 & 45 & 23 & 44 & 8.97 \\
M   & -0.99 & 41 & 57  & 8.5 & 8.97 \\
\tableline
\tableline
\end{tabular}
\end{center}

\tablenotetext{*}{SEP reduces to the one given by Nowak \& Wagoner (1991) for $ a = 0$.}

\tablenotetext{**}{The potential given by Mukhopadhyay (2002) 
reduces to the Paczy\'nski-Wiita potential for $ a = 0$.}

\end{table*}

\end{document}